\documentclass[
aps,
prd,
showpacs,
preprint,
tightenlines,
nofootinbib,
floatfix]
{revtex4}
\usepackage{graphicx,
}

\newcommand{\be}{\begin{equation}}
\newcommand{\ee}{\end{equation}}
\newcommand{\bea}{\begin{eqnarray}}
\newcommand{\eea}{\end{eqnarray}}

\newcommand{\ApJS}{{\it Astrophys. J. Suppl.\,}}

\newcommand{\etal}{{\it et al.}}
\def\ibid#1#2#3{{\it ibid. }{\bf #1},~#3~(#2)}

\def\fun#1#2{\lower3.6pt\vbox{\baselineskip0pt\lineskip.9pt
        \ialign{$\mathsurround=0pt#1\hfill##\hfil$\crcr#2\crcr\sim\crcr}}}

\newcommand\lsim{\mathrel{\rlap{\lower4pt\hbox{\hskip1pt$\sim$}}
    \raise1pt\hbox{$<$}}}
\newcommand\gsim{\mathrel{\rlap{\lower4pt\hbox{\hskip1pt$\sim$}}
    \raise1pt\hbox{$>$}}}

\def\dslash{\not{\hbox{\kern-2pt $\partial$}}}
\def\Dslash{\not{\hbox{\kern-4pt $D$}}}
\def\Oslash{\not{\hbox{\kern-4pt $O$}}}
\def\Qslash{\not{\hbox{\kern-4pt $Q$}}}
\def\pslash{\not{\hbox{\kern-2.3pt $p$}}}
\def\kslash{\not{\hbox{\kern-2.3pt $k$}}}
\def\qslash{\not{\hbox{\kern-2.3pt $q$}}}

 \newtoks\slashfraction
 \slashfraction={.13}
 \def\slash#1{\setbox0\hbox{$ #1 $}
 \setbox0\hbox to \the\slashfraction\wd0{\hss \box0}/\box0 }

\def\ee{\end{equation}}
\def\be{\begin{equation}}

\begin{document}

\setlength{\unitlength}{1mm}
\title{Constraints on Dark Energy and Distance Duality 
from Sunyaev Zel'dovich Effect and
Chandra X-ray measurements}

\author{Francesco De Bernardis,
Elena Giusarma and Alessandro Melchiorri}
\address{Physics Department and sezione INFN, University of Rome ``La Sapienza'',
Ple Aldo Moro 2, 00185 Rome, Italy\\
}
\date{\today}%

\begin{abstract}
We demonstrate that the recent measurements of the angular diameter
distance of $38$ cluster of galaxies using Chandra X-ray data and
radio observations from the OVRO and BIMA interferometric arrays 
place new and independent constraints on the dark energy
component. In particular we found that the equation of state 
is bound to be $-1.18 < w <-0.35$ at $68\%$ c.l..
We also search for deviations in the duality relation
between angular and luminosity distances. 
Using only cluster data, we found that the ratio between 
the two distances defined 
as $\eta = D_L/D_A(1+z)^2.$ is bound to be $\eta=0.97\pm0.03$ at 
$68 \%$ c.l. with no evidence for distance duality violation
in the framework of the $\Lambda$-CDM model.
Comparing the cluster angular diameter distance data 
with luminosity distance data from type Ia Supernovae, 
we obtain the
model independent constraint $\eta=1.01\pm0.07$ at $68 \%$ c.l..
Those results provide an useful check 
for the cosmological concordance model and for the
 presence of systematics in SN-Ia and cluster data.
\end{abstract}
\bigskip

\maketitle

\section{Introduction} 

One of the most important question in modern cosmology is
to understand the nature of the Dark Energy component 
of our Universe. Recent cosmological data coming from
measurements of the Cosmic Microwave Background (CMB) anisotropies
(see e.g. \cite{Bennett03}), on galaxy clustering (see e.g.
\cite{Tg04}) and, more recently, on Lyman-alpha Forest clouds (see
e.g. \cite{Se04}) are indeed in  spectacular agreement with the
expectations of a cosmological model where about $70 \%$ of
its energy density is in the form of a dark, unclustered, 
component (\cite{Se04},\cite{spergel}).
When further combined with measurements of luminosity distance
$D_L(z)$ of high redshift type Ia supernovae (SN-Ia, see \cite{riess}) 
the data provide compelling evidence that the universe is
currently undergoing an acceleration phase, i.e. the equation 
of state of the dark energy component must be 
$P/\rho = w < -1/2$.
Several theoretical models have been proposed to 
explain dark energy but none of them seems to provide 
a simple and natural solution to the
cosmological constant problem (why is the dark energy density
so small?) and to the ``Why now?'' problem (why the dark energy
dominates today ?) (see e.g. \cite{cosmo} and references therein). 
Moreover since systematics may be present in SN-Ia data, 
is crucial to test the apparent late time acceleration 
using the largest amount of independent and complementary 
information.

\noindent One possible way to test the results from 
high reshift SN-Ia is to measure the angular diameter 
distance $D_A(z)$ of high redshift objects and 
verify the distance duality relation:

\be
D_L(z)=(1+z)^2D_A(z)
\ee

Any systematic (both experimental and theoretical)
in the determination of the luminosity distance and 
in its interpretation will indeed break this relation (see e.g. 
\cite{kunz}).
Unfortunately a measurement of $D_A$ is a very 
difficult task. One possible way is to use 
FRIIb radio galaxies or compact radio sources but these
methods may be plagued by several systematics.
The most promising way to measure the angular distance 
is using the Sunyaev-Zel'dovich (SZ) effect togheter with 
X-ray emission of galaxy clusters.
Briefly, by combining the Cosmic Microwave Background 
 temperature decrement due to the SZ effect with measurements
of the X-ray surface brightness, one can in principle
determine the typical size of the line of sight inside the 
cluster and measure its angular diameter distance.
In this respect, it is particular timely to analyze the 
latest results on the cosmic distance scale from 
X-ray data and SZ effect measurements of high redshift clusters 
presented by Bonamente et al. 2006 (\cite{bonamente}).
In \cite{bonamente} the angular diameter distance of $39$ clusters in 
the redshift range $0.14<z<0.89$ has been determined
using Chandra X-ray data and radio observations from the 
Owens Valley Radio Observatory (OVRO) and 
Berkeley-Illinois-Maryland Association (BIMA) interferometric arrays.
The goal of our paper is therefore to make use of this new
high-quality data in order to constrain properties of dark energy 
and violations of the distance duality relation. 
We do this following $3$ steps:
first of all, we compare the new angular distance data
with a database of theoretical models, providing new
constraint on the dark energy equation of state parameter $w$.
Secondly, we investigate how the new data is compatible
with the current ``concordance'' cosmological model
constraining the $\eta$ parameter mentioned above.
Finally we compare the angular distance data with
the latest luminosity distance observations of SN-Ia
further constraining possible systematics in both datasets 
and investigating signatures of new physics.
In the next section we describe our method analysis and
our results. In the final section we 
derive our conclusions.

\section{Data Analysis and Results}

The data we use in this paper comes from the 
results presented in \cite{bonamente}.
We consider the angular diameter distance data of $38$ 
clusters presented in Table 2
of \cite{bonamente} and we associate to each measurement an error
$\sigma_{Clusters}$ which is obtained by combining the uncertainties 
in the cluster modelling 
plus the statistical and systematic errors.
The contribution to the statistical errors comes
primarly from the cluster asphericity, SZ point sources and the
kinetic SZ effect. Statistical errors affect the measurement of
$D_A$ at the level of about $\sim 20 \%$. The systematic contributions come
from the X-ray absolute flux calibration, the X-ray temperature
calibration and the SZ calibration. The overall effect from
systematics is about $\sim 15 \%$ of the measured signal. 
The data is then compared with the prediction of the angular
diameter distance in a flat universe given by (in units
of $c=1$):

\be 
D_A(z)={H_0^{-1} \over (1+z)}\int_0^z {dz' \over {E(z')}}
\ee

\noindent where $H_0$ is the Hubble constant and

\be
E(z)=((1-\Omega_w)(1+z)^3+\Omega_w(1+z)^{3(1+w)})^{1/2}
\ee

\noindent where $\Omega_w$ is the 
dark energy density component in units of the critical
density $\rho_c=3H_0^2/8 \pi G$.
For each theoretical model we
 then evaluate the likelihood function $exp(-\chi^2/2)$ where 

\be
\chi^2=\sum_{i} {({{D_A^{Cluster}(z_i)}-D_A(z_i)})^2 \over \sigma^2_{Clusters,i}}
\ee

\noindent with $z_i$ as the redshift of the $i-th$ cluster
and $D_A^{Cluster}$ is its measured angular diameter distance.
Fixing $\Omega_m=0.27$ and $w=-1$ we constrain the Hubble parameter 
as $H_0={76.8_{-2.9}^{+3.7}}_{-8}^{+10}$ 
(first error statistical, second systematic) in
extremely good agreement with the results presented
in \cite{bonamente}.

In Figure 1 we report the constraints obtained on the
$H_0-w$ plane using the cluster data. Since the cluster 
data alone is not powerful enough in determining the 
dark energy component we combine the cluster data with the HST result on
the Hubble parameter $h=0.72\pm0.07$ at $68 \% c.l.$ 
(see \cite{freedman} and we also include a gaussian prior on the lower value of 
the age of the universe as $t_0=12\pm1 Gyrs$ when the age of 
the theoretical model is $t<t_0$. We also restrict the
analysis to $\Omega_m=0.3$.
As we can see, this analysis bounds $w$ to be 
$-1.18<w<-0.35$ at $68 \%$ c.l.. This constrain is
less stringent than the one obtained from recent combined 
analysis of cosmic microwave background anisotropies, SN-Ia 
and large scale structure data (see e.g. \cite{spergel}).
It is however completely independent, constrains in a stronger 
way {\it phantom} $w <-1$ models 
and provides an useful cross-check of the current cosmological 
concordance model.

\begin{figure}[t]
\begin{center}
\includegraphics[width=1.0\linewidth]{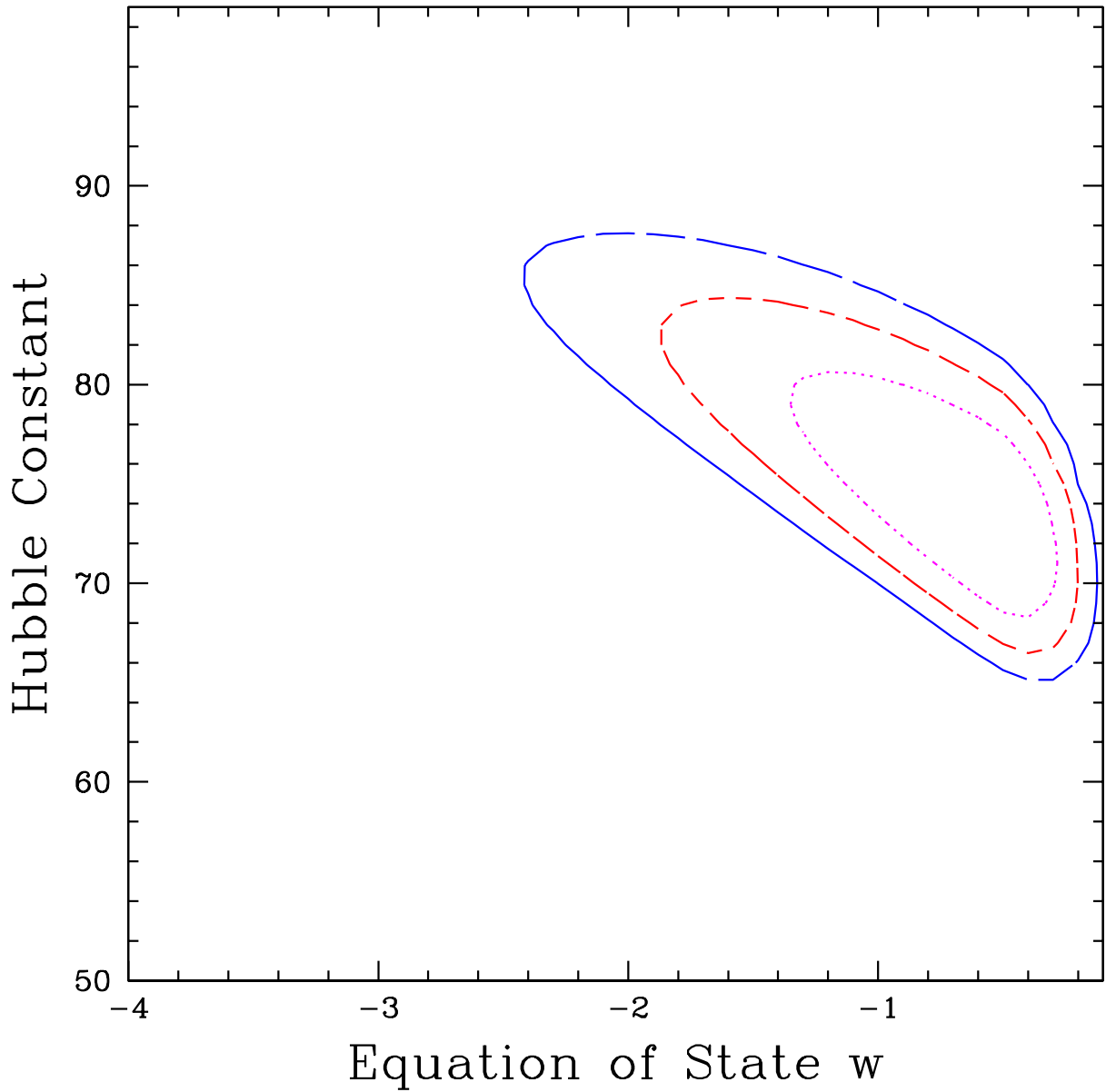}
\caption{Constraints on the $H_0-w$ plane from current
angular distance measurements of high redshift clusters.
The dotted,slash and solid lines correspond to $68 \%$,
$95 \%$, $99 \%$ confidence levels respectively.
Flatness and priors on the Hubble constant and on the age of 
the Universe are assumed (see text).
} 
\end{center}
\end{figure}

\begin{figure}[t]
\begin{center}
\includegraphics[width=1.0\linewidth]{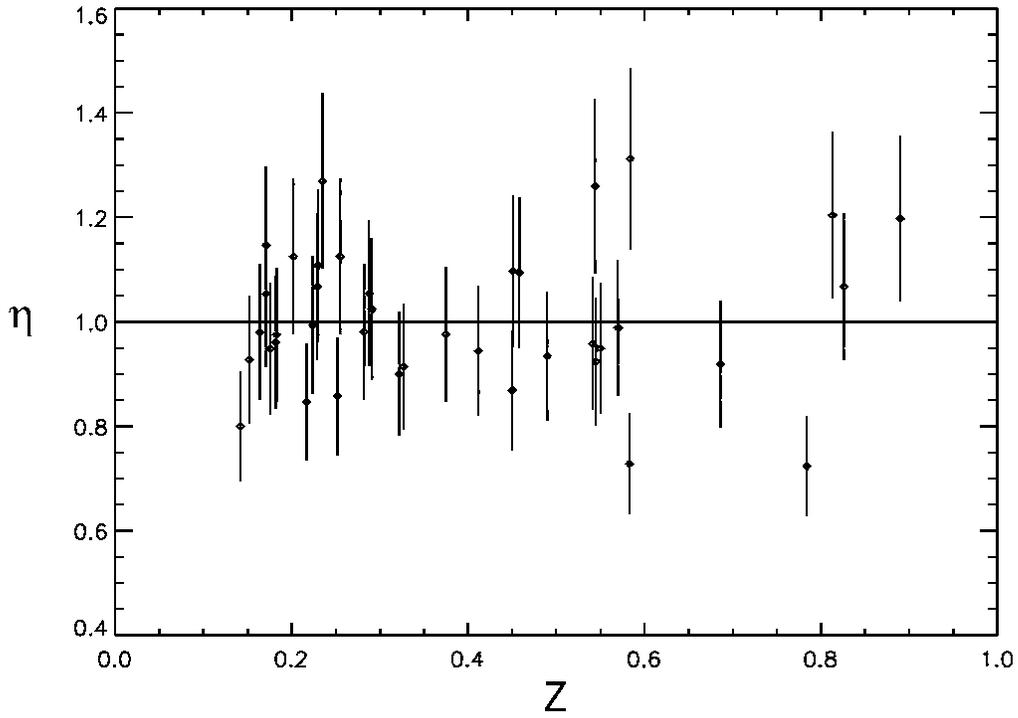}
\caption{Values of  $\eta=\sqrt{D_A^{Cluster}/D_A^{Theory}}$ 
derived from each cluster angular diameter distance and the
expected angular distance in the concordance model. The error
bars include the statistical errors on each cluster plus the 
uncertainty in the cosmological concordance
model. The data is in agreement with a constant-with-redshift
$\eta=1$ with  $\eta =0.97 \pm 0.03$ at $68 \%$ c.l..} 
\end{center}
\end{figure}

\begin{figure}[t]
\begin{center}
\includegraphics[angle=-90,width=1.05\linewidth]{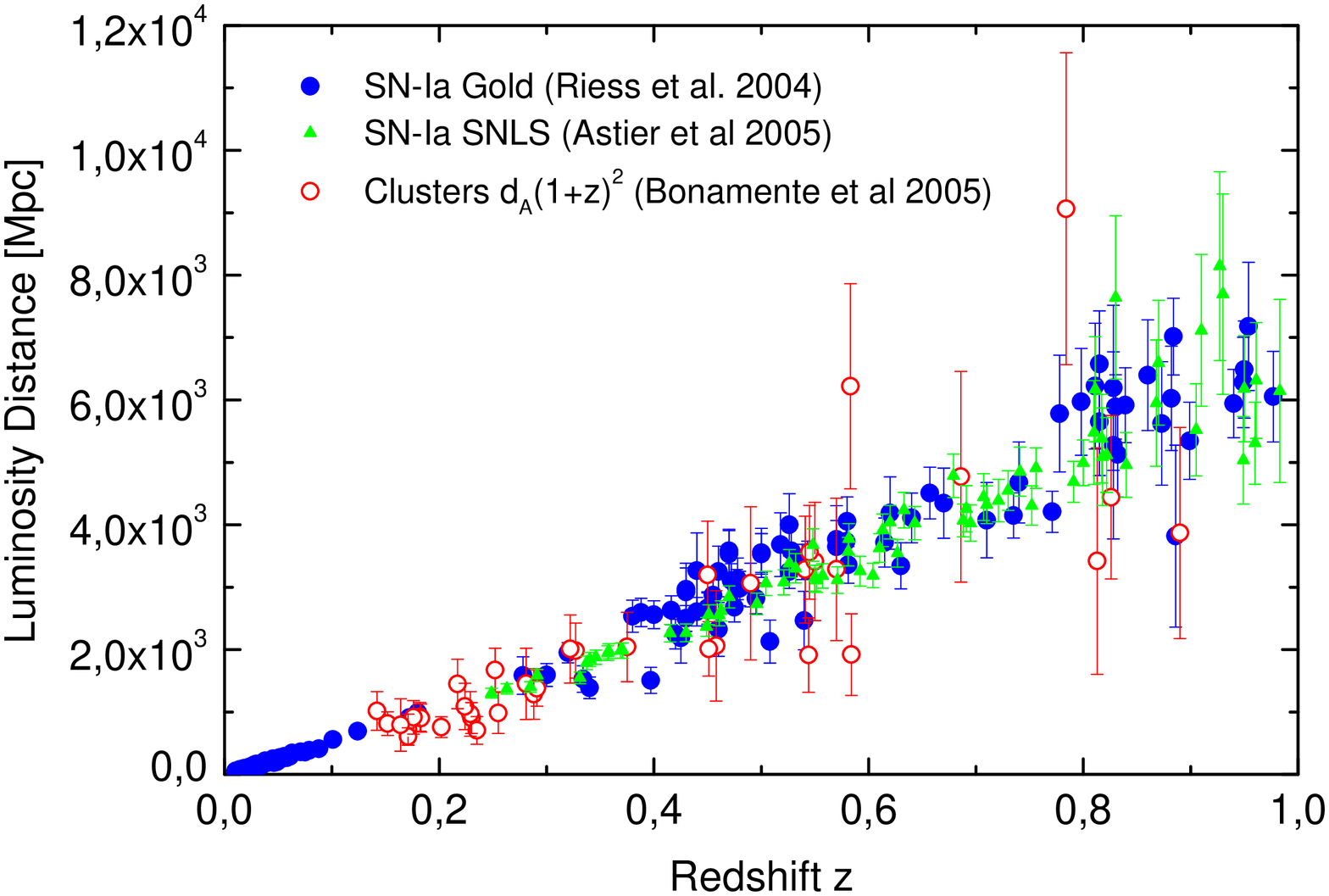}
\caption{Comparison of luminosity distance data from SN-Ia (Riess et
  al. 2004, Astier et al. 2005)) and angular distance data 
from SZ/X-Ray cluster
  observations (rescaled by $(1+z)^2$). 
The datasets are compatible providing no
indication for systematics and/or modification of the duality
distance relation.}
\end{center}
\end{figure}

As next step we check the consistency of the angular diameter distance 
cluster data with the current theoretical concordance model
by computing a database of $\Lambda$-CDM models with
$\Omega_m=0.30\pm0.10$, $h=0.73 \pm 0.04$,  
$\Omega_{\Lambda}=0.7\pm0.01$. As pointed out by \cite{uzan}, measuring 
the X-ray surface brightness involves a measurement of
the luminosity distance as well. Therefore if the
distance duality relation {\it is} violated:

\be
\eta(z)={{D_L(z)}\over{D_A(1+z)^2}}\neq1
\ee

then the angular diameter distance measured by the cluster
will be (see \cite{uzan}):

\be 
D_A^{Clusters}(z)=D_A(z)\eta^2(z)
\ee

Before a comparison  with luminosity distance data
we have therefore to compare the clusters angular distance 
 with the theoretical expectations of the standard $\Lambda-CDM$ 
model in order to see if the data is consistent with no 
violation of the distance duality relation (see \cite{uzan}).

We then aestimate the values of the parameter:

\be
\eta(z)=\sqrt{D_A^{Cluster}/D_A^{Theory}}
\ee

\noindent where the error bars on this quantity are aestimated
by combining the experimental error bars and the uncertainties
on the concordance model as in \cite{uzan}.

In Fig.2 we plot the results of this analysis.
As we can see, the data is in good agreement with a 
constant value of $\eta=1.0$, i.e. yielding no indication
for a violation of the distance-duality relation.
Considering $\eta$ as a constant but varying its amplitude we
obtain $\eta =0.97 \pm 0.03$ at $68 \%$ c.l. with a best-fit 
of $\chi^2=32.1$ with $38$ clusters.
This result should be compared with the previous 
result of $\eta=0.91\pm0.04$ by \cite{uzan}
based on $17$ clusters from the catalog of \cite{reese}.
Our analysis of the new data therefore improves previous constraints 
and is more consistent with no violation
of the distance duality relation

As final step, we compare the cluster data with the
luminosity distance of high redshift type Ia supernovae
from \cite{riess} and \cite{astier}. Since no violation of 
the distance duality
is observed in the cluster data we assume that the cluster
data provide a faithful aestimation of the angular distance.

If systematics are present in the SN-Ia, like non conservation
of the photon number by absortion from an unknow 
dust component, we can parametrize their effect as:

\be
D_L^{SN-Ia}(z)=\eta(z) D_L(z)
\ee

\noindent  yielding:

\be
D_L^{SN-Ia}={\eta (z)} D_A(1+z)^2
\ee

In Figure 3 we plot the recent luminosity distance 
SN-Ia data with the angular diameter
distance data multiplied by $(1+z)^2$. As we can see,
already from a first qualitative analysis, the datasets are 
consistent yielding no strong indication for a violation
of the distance duality relation.
In order to compare the datasets in a quantitative way we consider 
the weighted average of the data in $7$ bins spanning the range
$z=0.15,...,0.8$. For each bin $i$ we then consider a possible variation
of the distance duality relation by a term $\eta_i$ assumed as
constant. For each value of $\eta_i$ we
 can therefore construct a likelihood distribution function
$e^{-\chi^2/}$ where:

\be
\chi^2=(d_L^i-\eta_i(1+z_i)^2d_A^i)^2/({\sigma_L^i}^2+\eta^2
(1+z)^4{\sigma_A^i}^2)
\ee

wher $d_L^i$ and $d_A^i$ are the weighted averages of the angular and
luminosity distances inside the bin (with error bars $\sigma_i^{L,A}$).

\noindent We plot in Figure 4 (Top Panel) 
the normalized likelihood distribution
functions for each $\eta_i$. As we can see a value of $\eta_i=1$ is 
consistent at $68 \%$ c.l. with the data in each bin, yielding no
evidence for a variation of redshift for $\eta$. 
Assuming $\eta$ as constant over all the redshift range and
combining all the datasets we obtain:

\be
\eta=1.01 \pm 0.07
\ee

\noindent at $68 \%$ c.l., i.e. yielding no evidence for variations
in the duality distance relation. The likelihood distribution function
for $\eta$ is plotted in Figure 4 (bottom panel).

\begin{figure}[t]
\begin{center}
\includegraphics[angle=-90,width=0.9\linewidth]{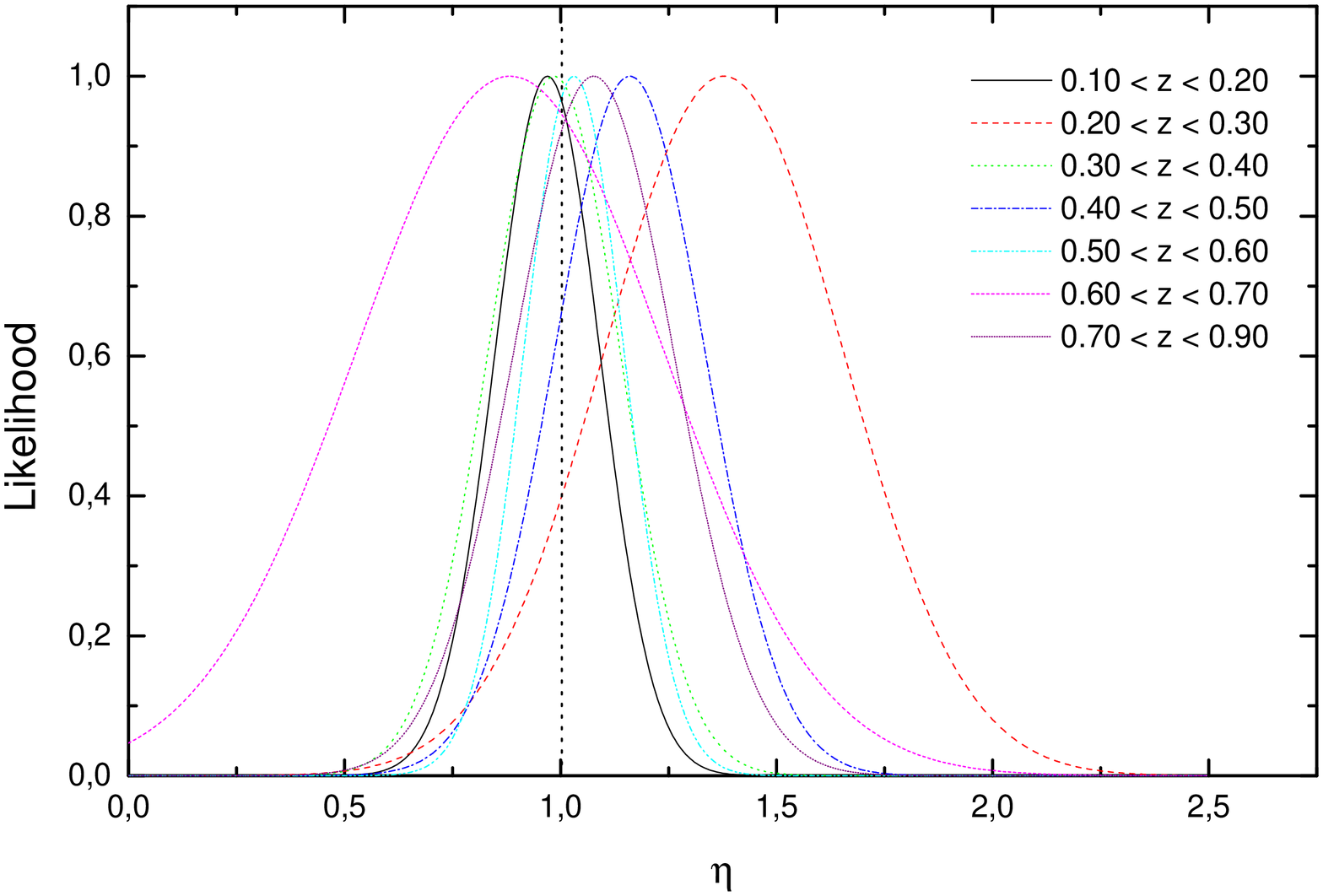}
\includegraphics[angle=-90,width=0.9\linewidth]{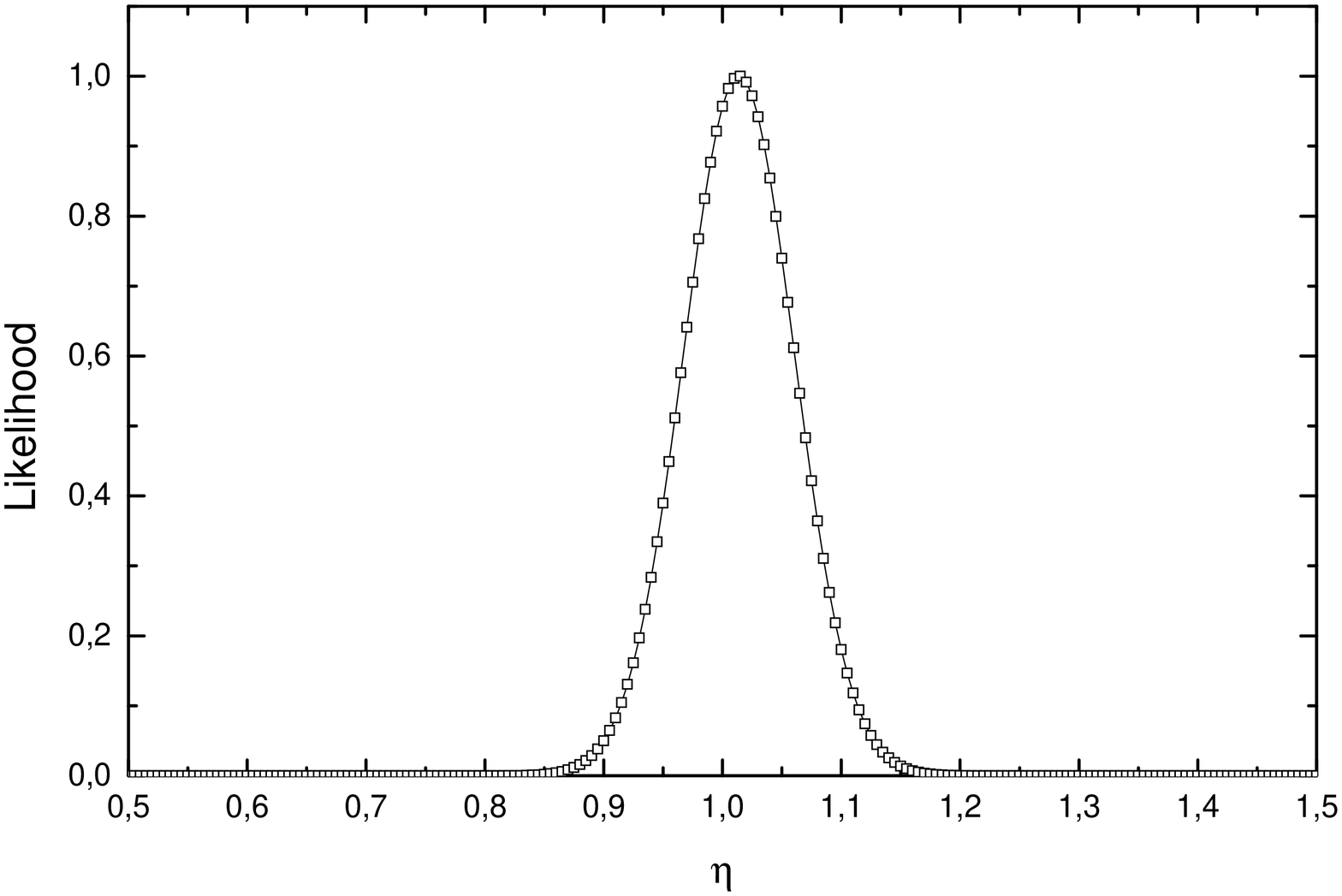}
\caption{Likelihood distribution functions for the values of 
$\eta_i$ in each bin (Top Panel) and for $\eta$ considered 
as constant over all
the redshift range spanned by the data (Bottom Panel). 
The data is consistent with
no violation of the distance duality relation ($\eta_i=1$)
and $\eta=1.01\pm0.07$ at $68 \%$ c.l..
}
\end{center}
\end{figure}

\section{Conclusions}

In this paper we demonstrated that the recent measurements of 
the angular diameter distance of $38$ cluster of galaxies 
using Chandra X-ray data and
radio observations from the OVRO and BIMA interferometric arrays 
place new and independent constraints on the dark energy
component. In particular we found that the equation of state 
is bound to be $-1.18 < w <-0.35$ at $68\%$ c.l.. 
We have also constrained possible
deviations from the duality relation between luminosity and angular
diameter distance. Those deviations may hint for systematics
like photon absortion by an unknown dust cmponent
or even be a signature of new physics like photon-axion
oscillation in an external magnetic field (see e.g. \cite{kunz}). 
We found that the ratio between the $2$ distances defined 
as $\eta = D_L/D_A(1+z)^2.$ is bound to be $\eta=0.97\pm0.03$ at 
$68 \%$ c.l. with no evidence for distance duality violation
in the framework of the $\Lambda$-CDM model.
We finally compare the cluster angular diameter distance data 
with luminosity distance data from SN-Ia obtaining  
the model independent constraint 
$\eta=1.01\pm0.07$ at $68 \% c.l.$. 
Those results provide an useful check 
for the cosmological concordance model and for the
 presence of systematics in SN-Ia and cluster data.
Future cluster data will reduce the effect of systematics,
provide a much better angular distance estimates and,
togheter with SN-Ia data, a stronger test of the duality
distance relation.

\textit{Acknowledgements}
AM is supported by MURST through COFIN contract no. 2004027755.


\begin{thebibliography}{99}


\bibitem{Bennett03} C.L. Bennett \etal, astro-ph/0302207

\bibitem{Tg04}  SDSS Collaboration,
                M.~Tegmark {\it et al.},
                Phys.\ Rev.\ D {\bf 69}, 103501 (2004).

\bibitem{Se04}  U.~Seljak {\it et al.},
                astro-ph/0407372.

\bibitem{spergel} D.~N.~Spergel {\it et al.},astro-ph/0302209.



\bibitem{wmap1}
C.L. Bennett {\it et al.}, \ApJS{148}{2003}{1};
G. Hinshaw {\it et al.}, \ibid{148}{2003}{135};

\bibitem{freedman}
W.~L.~Freedman {\it et al.},
Astrophys.\ J.\  {\bf 553} (2001) 47.

\bibitem{uzan}
  J.~P.~Uzan, N.~Aghanim and Y.~Mellier,
  Phys.\ Rev.\ D {\bf 70}, 083533 (2004)
  [arXiv:astro-ph/0405620].

\bibitem{kunz}
  B.~A.~Bassett and M.~Kunz,
  Phys.\ Rev.\ D {\bf 69}, 101305 (2004)
  [arXiv:astro-ph/0312443].

\bibitem{bonamente}
  M.~Bonamente, M.~K.~Joy, S.~J.~La Roque, J.~E.~Carlstrom, E.~D.~Reese and K.~S.~Dawson,
  arXiv:astro-ph/0512349.

\bibitem{reese}
  E.~D.~Reese, J.~E.~Carlstrom, M.~Joy, J.~J.~Mohr, L.~Grego and W.~L.~Holzapfel,
  Astrophys.\ J.\  {\bf 581} (2002) 53
  [arXiv:astro-ph/0205350].


\bibitem{riess}
  A.~G.~Riess {\it et al.}  [Supernova Search Team Collaboration],
  Astrophys.\ J.\  {\bf 607}, 665 (2004)
  [arXiv:astro-ph/0402512].

\bibitem{cosmo}
A.~Melchiorri, L.~Mersini-Houghton, C.~J.~Odman and M.~Trodden,
  Phys.\ Rev.\ D {\bf 68} (2003) 043509
  [arXiv:astro-ph/0211522];
R.~Bean and A.~Melchiorri,
  Phys.\ Rev.\ D {\bf 65} (2002) 041302
  [arXiv:astro-ph/0110472].


\bibitem{astier}
  P.~Astier {\it et al.},
  arXiv:astro-ph/0510447.


\end{thebibliography}
\end{document}